# FABRICATION AND PROPERTIES OF GALLIUM METALLIC PHOTONIC CRYSTALS

V.F. Kozhevnikov, M. Diwekar, V.P. Kamaev, J. Shi, and Z.V. Vardeny
Physics Department, University of Utah, Salt Lake City, UT 84112

Gallium metallic photonic crystals with 100% filling factor have been fabricated via infiltration of liquid gallium into opals of 300-nm silica spheres using a novel high pressure-high temperature technique. The electrical resistance of the Ga-opal crystals was measured at temperatures from 10 to 280 K. The data obtained show that Ga-opal crystals are metallic network with slightly smaller temperature coefficient of resistivity than that for bulk gallium. Optical reflectivity of bulk gallium, plain opal and several Ga-opal crystals were measured at photon energies from 0.3 to 6 eV. A pronounced photonic stop band in the visible spectral range was found in both the plain and Ga infiltrated opals. The reflectivity spectra also show increase in reflectivity below 0.6 eV; which we interpret as a significantly lower effective plasma frequency of the metallic mesh in the infiltrated opal compare to the plasma frequency in the pure metal.

## 1. INTRODUCTION

In recent years there have been fast growing interest in photonic crystals (PC), which are materials with periodically modulated dielectric constant [1, 2]. It was predicted theoretically [3-8] that such structures possess a band gap for electromagnetic waves similar to that for valence electrons in crystalline semiconductors. Specific interest is associated with periodical metallic mesh or metallic photonic crystals (MPC). Theoretical calculations suggest that the MPC have a larger band gap than that in pure dielectric structures and support propagation of electromagnetic waves at frequencies lower than the plasma frequency of the underlying metal they are made of [9-11]. There is a common belief that PC will find a wide spectrum of applications in opto-elecronics [1,2]. One of them is fabrication of "left handed" materials on the basis of MPC made of ferromagnetic metals [12,13].

The main challenge in the experimental studies of both metallic and dielectric PC is caused by difficulties in sample fabrication. These difficulties are more acute for large sample size and for PC with higher band gap. For band gap in the visible spectral range the PC lattice constant should be in the sub-micron dimension range. For this reason the majority of the experimental studies on PC have been completed for PC with band gap in the microwave to millimeter wavelength ranges [14-16]. Currently the most promoted approach for fabricating three dimensional (3D) MPC for the visible spectral range is based on infiltration of metals into a colloidal PC that serves as a template. The colloidal PC is usually fabricated of monodispersed $SiO_2$ (silica) spheres [17-19]. This approach is used in the present work where the PC template is an artificial opal formed of 300-nm silica spheres.

We can distinguish three major problems in fabricating 3D MPC based on opal templates.
(1) Fabrication of opals with low defect density. Defects in PC create localize states within the photonic band gap similar to that in crystalline semiconductors. In addition, dispersion in size and shape of the silica spheres may smear the optical gap and forms band tails similar to that in amorphous semiconductors. Our opals contain a sizable defect density that limits the light mean free path to 5 µm, as recently determined using coherent backscattering [20].
(2) The metal infiltration should not create new structural defects. The main culprit of metal infiltrated opals is the filling factor: a ratio of the volumes of infiltrated metal and all the pours within the opal. Filling factor less than 100% implies empty space in voids and braked contacts between the voids. Such defects are randomly distributed in the MPC end product and may dramatically increase the total defect concentration. We note that filling by liquid metal poses fewer difficulties with the filling factor, and this is the main reason for choosing gallium in this work, since it is a liquid close to room temperature. In addition, gallium, being a simple metal is one of a few substances that *expanses* at solidification and this also helps with the filling.
(3) Three basic techniques are currently used for metal infiltration: chemical vapor deposition, electroplating, and aforementioned liquid metal infiltration under high pressure. We have developed a new modification of the high-pressure technique that allows us to reach complete filling of the opals. But for this technique (as well as for the traditional high-pressure filling [18,19]) one faces the problem of nondestructive extraction of the metal infiltrated opal from the surrounding solid metallic "bullet". Mechanical cutting with a standard, such as Buehler Isomet type diamond saw, significantly destroys the infiltrated opals and renders them unsuitable for optical measurements. For this reason we used only originally grown opals having at least one smooth surface. The difference in the thermal expansion of bulk gallium and gallium infiltrated opals allowed us to nondestructively extract the opal MPC from the gallium bullet.

In this paper we report preliminary results on gallium infiltrated MPC with 100% filling factor. We report electrical and optical measurements that prove the high filling factor and hint to the possibility of a low frequency plasma edge. A novel high pressure-high temperature filling technology used for the sample fabrication is described in the next section. The data on electrical transport and optical reflectivity are presented in the following two sections. The last section contains a summary.

## 2. SAMPLE FABRICATION

The opals used as template for the metal infiltration have been grown via gravity sedimentation of 300-nm silica balls (Sicastar plain, Micromod Partikeltechnologie GmbH) from 50 mg/ml aqueous suspension. After about a two-week sedimentation process at room temperature, and consequent drying at 150° C, and baking at 950° C, opal PC of about 0.7 mm thick were formed. The opal samples had one shiny and one mat surface.

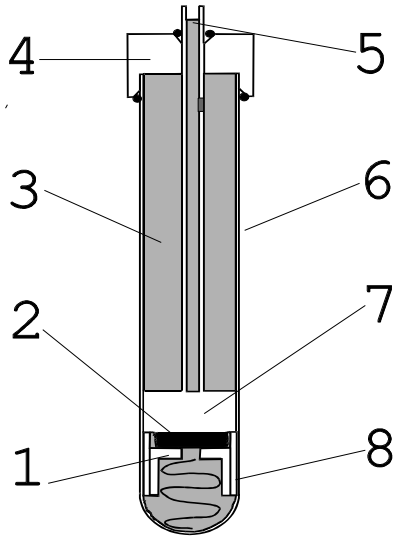

Fig.1. The quartz ampoule with opal.
1, supporting cup with soft spring; 2, opal; 3, gallium; 4, plug (brass); 5, tube for pumping out and filling the ampoule; 6, quartz ampoule; 7, stopper; 8, quartz tube.

A relatively small chip of the opal (from 4 to 7 mm in size) was placed in a quartz ampoule (95 mm length, 9 mm inner diameter and 1 mm wall) between upper stopper and lower cup as shown in Fig.1. Both the stopper and cup were made of stainless steel. The ampoule was pumped out to pressure of the order of $10^{-6}$ torr, and then completely filled by liquid gallium (99.9999% purity) at temperature about 35-40° C (the melting temperature of gallium is 30° C). The ampoule was installed in a homemade high-pressure vessel (thermal treated steel). The lower part of the ampoule was within a 1.5 kW heater made of a nickel-chromium wire. The working pressure of the high-pressure vessel was up to 2 kbar. The vessel was attached to a homemade gas compressor with an air driving oil pump and piston multiplier. Tungsten-rhenium thermocouples were used to control the ampoule temperature.

The hermetically sealed vessel was first filled with argon at pressure of 10 bar. Then the sample was heated up to 1000 K and kept at this temperature for about 3-5 minutes. At the same time some argon gas at pressure of 2.2 kbar was collected in the compressor. Then the valve between the vessel and compressor was open wide to allow the compressed argon gas to fill the vessel. The argon gas pressure in the vessel at high temperature was about 600 bar. Then the heater was turned off and the sample was cooled down to temperature of the cooling waters (10-14° C). Then the pressure was released and the vessel was disassembled. The time interval during which the pressure was applied to the liquid gallium was about 1 minute. We have intentionally limited this exposure time to prevent an effect of solubility of argon in the liquid gallium that may have lead to penetration of argon into the sample through the gallium column. An estimate for pressure needed in order to infiltrate liquid gallium through the opal pours is 700 bar. For this estimate we used the Laplace formula for the pressure drop in a capillary with 700 mN/m surface tension of gallium [21] and 20 nm radius of the necks that connect the voids between the silica spheres in the opal PC. Although the estimated pressure is likely to be an upper limit of the Laplace pressure in the necks, it is also possible that the observed complete filling of the opals occurs due to a wetting transition on the gallium-silica interface at temperature somewhere below 1000 K.

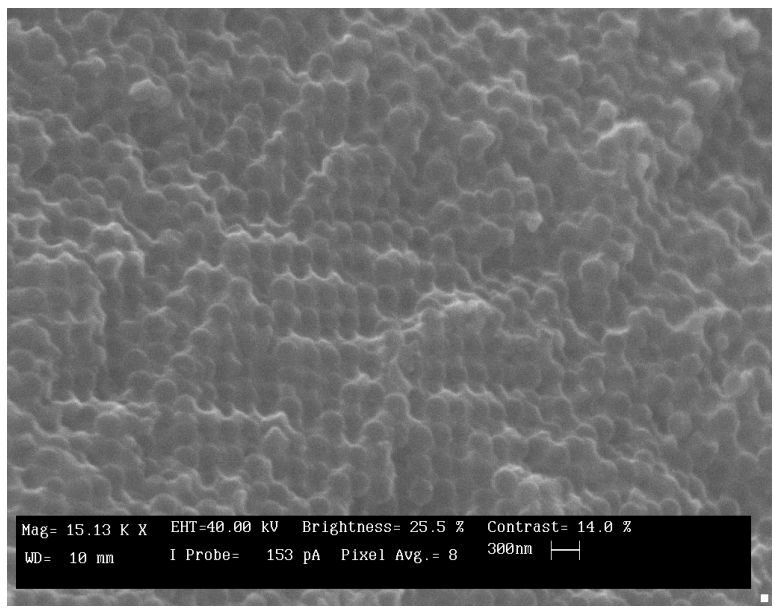

Fig.2. ESM image of a cleaved edge of gallium infiltrated opal.

Liquid metal penetrated into the opal mostly through its parallel sides. The depth of penetration (thickness of the metal infiltrated layer) depends on wettability, viscosity and time. The initial vacuum guaranteed that both upper and lower infiltrated layers were completely filled by the liquid metal. Taking into account the abnormal change of gallium density at melting, no holes may appear in the layers at solidification. Imaging the cleaved edges of the sample under optical and electronic microscopes checked the quality of the infiltration (layer overlap) as seen in Fig. 2. All of our samples have had uniformly filled cleaved edges of a dark green-blue color. Thus we conclude that the obtained gallium infiltrated opals have 100% filling factor. As was mentioned above, the MPC samples were extracted from the bulk solid gallium using the difference in the thermal expansion of the gallium-infiltrated opal and bulk gallium. We succeeded in obtaining samples with the same shiny and mat surfaces as the original plain opal. Having a shiny surface for the metallic opal was important for the optical reflectivity measurements.

## 3. ELECTRICAL CONDUCTIVITY MEASUREMENTS

Data on the temperature dependence of electrical resistance $R/R_0$ for gallium-infiltrated opal (three sets) and bulk gallium are shown in Fig. 3. Here $R_0$, the sample resistances at 273 K, is 2.04 m$\Omega$ for the Ga opal, and 0.12 m$\Omega$ for the sample of bulk Ga. The measurements have been performed by a four-probe technique with 0.5 mA dc-current. From the linear dependence of R(T), which is typical for metals we conclude that the gallium infiltrated opal is a metal, and thus our samples represent true MPC. Other MPC that were fabricated elsewhere using opal templates, so far have not shown a linear increase of R with temperature and thus do not represent 100% filling. We conclude therefore that the gallium-infiltrated opals are the first true MPC samples.

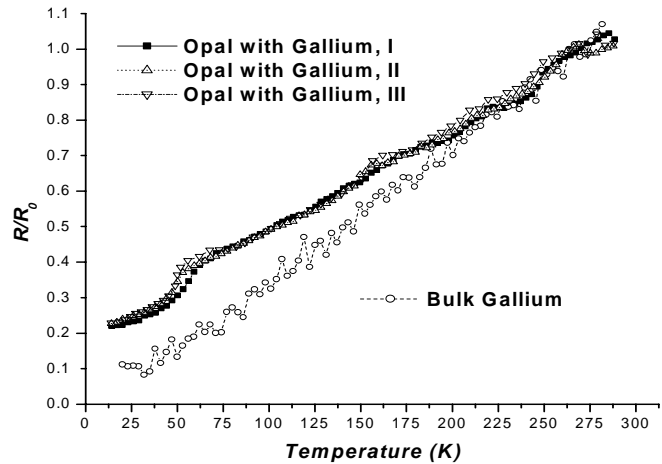

Fig.3. Electrical resistance of bulk gallium and gallium MPC.

From Fig. 3 we also conjecture that the temperature coefficient of resistivity for gallium infiltrated opal is smaller than that for the bulk metal. This is consistent with a known phenomenon of change of matter properties in porous materials (see, for example, Refs. 22, 23). In addition, it is seen that near 50 K R(T) of the gallium MPC exhibits a prominent change; actually R(T) shows an exponential drop at this temperature. This is consistent with a structural phase transition recently observed via X-ray diffraction [23]. No such change exists in bulk gallium. This very interesting confinement induced phenomenon deserves a special investigation.

## 4. OPTICAL REFLECTIVITY MEASUREMENTS

The optical reflectivity spectra of bulk gallium and gallium infiltrated opal (both at 70 K), and plain opal at room temperature are shown in Fig. 4. The spectra were measured using a home made spectrometer in the spectral range from 0.3 to 6.0 eV. The spectrometer was composed of a various combinations of diffraction gratings, optical filters and solid state detectors such as UV-enhanced silicon, germanium and InAs. The dispersion element was a ¼ met Jerrel-Ash mono-spectrometer that was operated using four different gratings. The pure gallium plasma frequency is about 15 eV and thus is outside the present optical capabilities of our spectrometer.

The reflectivity spectrum of the pure gallium film (Fig. 4(a)) is rather flat with a smooth decrease at around 3 eV, which may be caused by the beginning of the metallic plasma edge coupled with a sizable attenuation. The reflectivity spectrum of the plain (uninfiltrated) opal (Fig. 4(b)) is also rather flat except for a pronounced Bragg reflectivity stop band at about 520 nm. This is caused by Bragg diffraction from the (111) planes of the opal in air. We note that the opal growing direction is along [111] and thus even if the opal sample is not a single crystal, nevertheless the [111] direction is well defined due to the opal growing process. The long tail of the reflectivity stop band towards longer wavelengths may be caused by longer interplane distance, that is due to insufficient sedimentation or the known dispersion in the silica sphere diameters. The relatively low reflectivity indicates relatively strong light scattering from the opal surface.

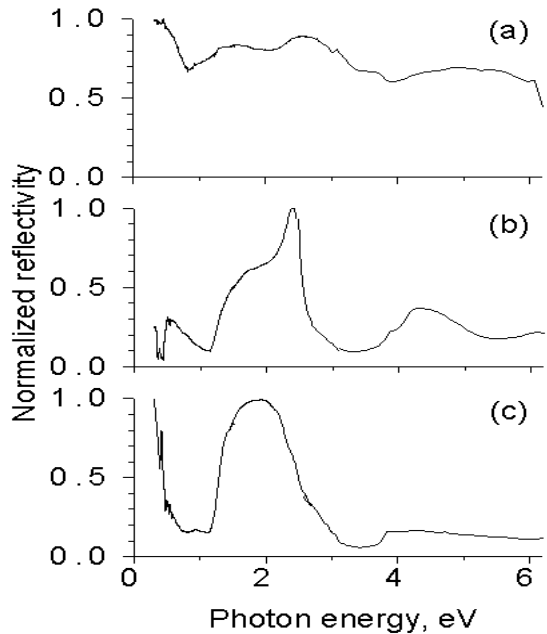

Fig.4. Optical reflectivity for bulk gallium (a); plain opal (b), and gallium infiltrated opal (c).

The reflectivity spectrum of the gallium MPC (Fig. 4(c)) is different from that of both plain gallium and opal samples. It contains a pronounced Bragg-like band gap with a broad peak at 650 nm (1.9 eV), but in addition it also shows an increase towards low energy. The Bragg-like stop band in the gallium MPC is red shifted compared to that of the plain opal because of the gallium metal filling. The effective refraction index of the gallium-silica combination in the MPC is probably larger than that of the air-silica combination in the plain opal and this causes the red shift of the MPC stop band. Obtaining a Bragg-like stop band in the MPC is already an achievement, since the skin depth of the gallium metal is about 10 nm, that is significantly less than one silica sphere layer. The appearance of the stop band in the reflectivity spectrum of the gallium infiltrated opal means that light penetrates in the MPC at least 10-15 (111) layers before it is scattered or absorbed. This indicates that the plasma frequency of the MPC cannot be higher than about 1.5 eV (the photon energy which corresponds to the Bragg stop band). We believe that the plasma frequency of the MPC is much lower than 1.5 eV. The reflectivity increase towards lower energies (Fig. 4c) may indicate that the actual plasma frequency of the MPC is less than 0.2 eV, as theoretically predicted. However more spectra, especially at lower photon energies are needed in order to prove that.

**5. SUMMARY**

Gallium-opal MPC with 100% filling factor have been fabricated for the first time. To achieve this, a new fabrication technique for the MPC was developed. The technique explores simultaneously high pressure and high temperature, and provides complete filling of opals with liquid metals. In case of gallium, it guarantees complete filling after the metal solidification. Using a difference in thermal expansion of gallium infiltrated opal and bulk gallium, the gallium MPC were extracted from the bulk gallium without destruction. In particular, the gallium-MPC have a shiny surface without the need of cutting and/or polishing of the sample after infiltration.

The measured temperature dependence of resistance for the gallium-infiltrated opals testifies that the obtained crystals are metallic. An interesting feature, namely a confinement induced phase transition in gallium near 50 K was revealed. The optical reflectivity spectrum of the gallium MPC was measured and compared with the reflectivity spectra of the uninfiltrated opal and a pure gallium film. Surprisingly the reflectivity spectrum of the MPC shows a Bragg-like stop band indicating a light penetration depth much larger than the pure gallium skin depth. We also observed the plasma edge of the MPC at much lower energies than the plasma edge of the pure gallium.

**ACKNOWLEDGEMENTS**

We thank Drs. S. Li and M. Delong for help with the opal fabrication, and Dr. M. Wohlgenannt for the help with the optical measurements. We also thank R. Anthon and D. Rodriguez for help with the sample preparation and diagnostic. Financial support of NSF (NIRT DMR–0102964) and the Army Research Office (DARP 19-00-1-0406) is greatly appreciated.